# Simulations of high intensity ions beam RFQ Cooler for DESIR/SPIRAL 2

**Ramzi Boussaid,**[a*] **Gilles Ban**[a,] **Jean Francois**[a,] **and Christophe Vandamme**[a]

[a] *LPC-IN2P3, ENSICAEN, 6 Boul. Maréchal Juin, 14050 Caen, France.*

*E-mail*: boussaidramzii@gmail.com
*ban@lpccaen.*in2p3.fr
cam@lpccaen.in2p3.fr
vandamme@lpccaen.in2p3.fr

ABSTRACT:

The transport simulation of a high intensity ions beam with a radio-frequency cooler within the SPIRAL 2/GANIL High Intensity Radiofrequency Cooler, SHIRAC, installed at the SPIRAL 2/DESIR facility, is presented. Two simulation methods of cooling process in presence of space charge effect will be studied. The beam properties degradation by the coulomb repulsion and the buffer gas diffusion will be discussed. Finally, a comparison in term of transmission between simulated and experimental results for $_{133}Cs^+$ ions beam of intensity going up to 1μA, will be outlined.

*Ramzi Boussaid, boussaidramzii@gmail.com and boussaid.ramzi2014@gmail.com

## Contents

## 1. Introduction

The next SPIRAL 2 facility held at GANIL laboratory (France) will enable us to study the very high intensity radioactive beams [1]. DESIR ('Decay Excitation and Storage of Radioactive Ions') is a low energy, up to 60 keV, beam facility. It will use a large variety of radioactive ions beams which will be available at SPIRAL2 [2]. As the ions beams will have an emittance around 80 $\pi$.mm.mrad [3], a high resolution separator (HRS) must be implemented in experimental line [4]. Then, This HRS nominal running will require a RFQ cooler [5, 6, 7].

In the context of this next-generation facility, ions beam intensities will reach a high intensity up to 1 µA which exceeds the current technology with intensity of few tens of nA [8, 9, 10, 11]. The cooling of these highest intensity beams requires a special consideration of the space charge effect. To counter this effect, highest RF amplitude of confinement and low inner radius of RFQ must be used [12, 13, 14]. Then, a new RFQ Cooler must be designed.

The HRS requirements give the following specifications for this RFQ Cooler: a transverse emittance ε < 3 π.mm.mrad at 60 keV, an energy spread ΔE of 1 eV and a transmission efficiency more than 20 % for ions with mass m>12 a.m.u, more than 40 % for m>40 a.m.u and more than 60 % for m>90 a.m.u [4].

This paper presents a simulation study of cooling of the highest intensities beam and their transport up to the extraction section exit.

## 2. Buffer gas RFQ Cooler

### 2.1 Background

A RFQ Cooler consists of a quadrupole, filled in by light buffer gas, on which an RF oscillating potential with opposite phase is applied. A detailed introduction to this device has been presented in several references [5, 15,16]. It provides a radial confinement of ions along the quadrupole. The ions full-motion can be composed of two components: the micro-motion and the macro-motion [5, 22]. The ions transversal motion is governed by the so-called Mathieu equations [17, 18, 19] and their stability is insured by the Mathieu parameter 0<q<0.908 which defined by the following formula [20, 21, 22]:

$$q = \frac{8eV_{RF}}{mr_0^2 w_{RF}^2} \quad (1)$$

Where: e, m, $V_{RF}$, $w_{RF}$ and $r_0$ are respectively, the elementary charged ion, the ion mass, the RF potential amplitude, the frequency of $V_{RF}$ and the quadrupole inner radius.

### 2.2 Necessity of RFQ Cooler

In the future SPIRAL2/DESIR facility, the rare used beams will be characterized by a wide cross-sectional size, high intensity and high radioactivity [3]. Then, the ions of interests are overwhelmed into the contaminants. In the other hand, we need an isobaric level purification of beams [4]. In this case, a high resolution separator (HRS) must be set-up between the ions source and the experimental room. This solution has been developed in CENBG laboratory located in Bordeaux (France) and the beams properties must be of ε < 3 π.mm.mrad and ΔE ~ 1 eV [4, 23]. Under such conditions, it is possible to achieve the isobaric purification level and reach a mass resolution of 30000 and a transmission close to 100 % [4, 23]. The only device which can provide the HRS requirements is a RFQ Cooler.

## 3 Main challenges

The experimental manipulation of high intensity beam requires dealing with three main physics effects: the space charge effect, the RF heating and the buffer gas diffusion.

### 3.1 Space charge considerations

The space charge means the ions-ions coulomb repulsion. Its effect on the cooled beam is the main specificity of this project. In this part, we will discuss its contribution in the degradation of the cooled beam properties.

#### 3.1.1 Space charge field: the tube method

At atmospheric pressure, it is very difficult to deal with the individual interaction between the ions, even with the fastest computers [24]. Few calculation methods are available, such as the tube-method [25], to give a quick and a rough estimate of the coulomb repulsion. In this method, the charge distribution is assumed as cylindrical and uniformly charged and the field is purely radial. Using the Gauss theorem, we obtain the following equations of field [26]:

- $r < r_s$ : $\overrightarrow{E_{sc}} = \frac{Ir}{2\pi\varepsilon_0 v r_s^2}\overrightarrow{u_r}$ (2.1)
- $r > r_s$ : $\overrightarrow{E_{sc}} = \frac{I}{2\pi\varepsilon_0 v r}\overrightarrow{u_r}$ (2.2)

Where $\varepsilon_0$ is the constant vacuum permittivity, v is the ion velocity, rs is the mean charge distribution radius, I is the beam current and r is a radial position.

These equations show direct competition between the current beam and the ions velocities; the slower the ions are, the more important the effect is. Along the cooling section, after the successive collisions with the buffer gas molecules, the ions will be slowed down from few hundred eV to just few eV [20]. Then, the space charge effect will be very important at the end of cooling section.

### 3.1.2 Space charge effect on the cooling

At the quasi-equilibrium state into the RFQ, the problematic of space-charge and its importance on the cooling is shown by the following equation [17, 27]:

$$m\frac{q^2}{8}w_{RF}^2 r_s^2 = 2k_B T + \frac{eI}{4\pi\varepsilon_0 v} \quad (3)$$

Where, the terms denote respectively the confinement term, the thermal term and the space charge term.

During the cooling, the thermal and confinement terms both decrease but the space charge term will increase due to its dependence on the ion velocity. Since, the only way to compensate this effect is to increase the RF voltage amplitude we have to study the RFQ Cooler confinement limit.

### 3.1.3 Space charge effect on the confinement capacity

In the Dehmelt concept, the RF voltage effect on the cooling is equivalent to a pseudo-potential well which traps the ions along the RFQ [28]. This well is assumed to be a harmonic oscillator one of depth D with a maximum allowable of [29]:

$$D_{max} = \frac{qV_{RF}}{4} \quad (4)$$

The space charge effect on the cooling is explained by the reduction of depth D of confinement potential [18]. To study quantitatively this reduction, the space charge potential is assumed to a harmonic oscillator well of depth DSC. This depth can be deduced from equation 2.1:

$$D_{SC} = \frac{I.r^2}{4\pi\varepsilon_0.v.r_0^2} \quad (5.1)$$

Its maximum is:

$$D_{SC,max} = \frac{I}{4\pi\varepsilon_0.v} \quad (5.2)$$

Then, the space charge effect is manifested by the reduction of confinement potential depth D by DSC [18]. Owing to this effect, the RFQ Cooler has a limit in the maximum intensity of the beam that can be cooled [30, 31]. The maximum intensity Imax is [32, 33, 34]:

$$I_{max} = \frac{2\varepsilon_0 D_{max}}{\pi.r_s^2.r_0^2.v} \quad (6)$$

For example: The needed RF parameters to cool a Cs+ ions beam, of intensity going up to 1µA and 1 eV of kinetic energy by an RFQ of 5 mm of inner radius, are a RF amplitude of 5.5 kV and a RF frequency of 4.5 MHz for a value of q=0.4.

### 3.2 Buffer gas diffusion

The buffer gas diffusion can occur through injection and extraction holes (see section 4). The vacuum system of a RFQ Cooler has the main challenge to keep the high vacuum in experimental beam line (10-5Pa) whilst adding a high gas load into the cooling section (up to a few Pa)[17, 35]. It consists of setting a high capacity differential pumping system [17, 36] and a smallest size of injection and extraction holes. These have diameters of 12 and 8 mm, respectively. On the below table, we have shown the pressure distribution of Helium buffer gas in different stages. Under such conditions, the pressures outside the RFQ chamber remain less than 1/200 of the RFQ pressure.

| Rate (ml/mn) | Injection pressure (Pa) | RFQ pressure (Pa) | Extraction pressure (Pa) |
|---|---|---|---|
| 15 | 8.19E-3 | 1.59 | 4.03E-3 |
| 20 | 1.22E-2 | 2.08 | 5.36E-3 |
| 25 | 1.68E-2 | 2.58 | 6.7E-3 |
| 30 | 2.1E-2 | 3.05 | 9.12E-3 |

**Table.1.** Measurement of buffer gas distribution pressure into the RFQ Cooler beam line

### 3.3 RF heating

The micro-motion of an ion colliding with an atom can make a phase jump with respect to the radiofrequency, which leads to an increase of the spread energy. Hence, the ion will be radially ejected out of the RFQ. This process is called RF heating [32, 37, 38]. Then, to avoid this process the q parameter has to be kept as low as possible ($q \leq 0.4$) [18, 5, 35] and a light buffer gas, in most cases Helium [17].

## 4 Numerical simulations tools

Our simulation attempts to study the space charge effect on the cooling process and transport of cooled beams in presence of buffer gas diffusion. This study is based on the simulation of the ion-gas collisions model.

### 4.1 Ion-gas collisions models

Several physics models involve simulating these collisions. In this paper we will quote two models: realistic potential model (RP) and hard sphere model (HS1).

#### 4.1.1 Hard sphere model HS1

The classical hard sphere collision model (HS) has been presented in details in these references [5, 39, 20]. Its execution is simple but it gives a bad agreement with experimental results [40] and it does not take into account the RF heating [2]. A first correction to this model is the first variable size hard sphere model, HS1 model [42, 43, 39, 5, 44]. This model showed us the decrease in the average kinetic energy of ions and their general decomposition movement toward the quadrupole field center. However, it cannot give quantitative accuracy on the cooling process at low ions energy, few eV [18].

#### 4.1.2 Realistic potential model

The realistic potential (RP) model is a not hard ion-gas potential. It is usually used to avoid the handicap of the HS1 model [12, 37, 45]. The mobility is calculated as a function of the collissional integral $\Omega(T_{eff})$ [5, 46, 47]:

$$K = \frac{3e}{16N} \sqrt{\frac{2\pi(m+M)}{K_B T_{eff} mM}} \frac{1}{\Omega(T_{eff})} \quad (7)$$

Where $K_B$ and $T_{eff}$ are the Boltzman constant and the effective temperature.

The RP model accuracy may be tested by comparing the calculated mobilities with the experimental mobility data [45]. The mobility in RP model is 1 to 11 % different to the experimental one, as you can see on the figure 1. This error seems to come from the experimental error and the error in the interaction potential form. This agreement test means the reliability modeling of the realistic interaction potential.

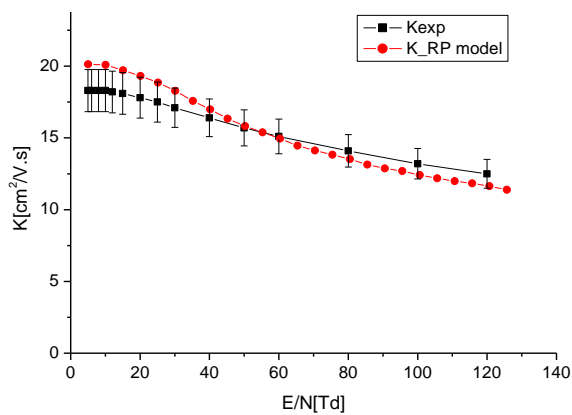

**Fig.1**: The experimental mobility data and the calculated mobility with the RP model for the $_{133}Cs^+$ ions in He gas at 300 K.

### 4.2 Simulations programs

In order to study the degradations of the cooled beam, two distinct modeling programs have been developed. The first one simulates the cooling process and the transport of the cooled beams by SIMION software and the other one simulates only the cooling process by Python program.

#### 4.2.1 Ion optics modeling program: SIMION 3D

SIMION is a ray tracing code for 3D simulation of the ions motions [25]. It provides the simulating of the ions motions in high pressure condition, under the influence of electrostatic field and in the presence of space charge effect [48]. The HS1 model has been implemented in this program with LUA language program [41]. SIMION program was adapted to the buffer gas diffusion.

SIMION incorporates two methods for estimating the space charge effect; the tube-method and the ion-cloud repulsion [41, 49]. The tube method is not available in high pressure because it is based on a space coherent assumption [41, 24]. The ion-cloud method assumes simply that the particles are charge clouds surrounding points. This method can be used in high pressure condition because it is based on a time coherent approach [41, 24].

The cooling process and transport ions beam can be achieved simultaneously by SIMION because it provides the distribution of electric field along the experimental line [41].

#### 4.2.2 Python program

To attempt to predict the ions motions in a neutral gas at high pressure, under the influence of electric field and in the presence of space charge effect, we have chosen to incorporate the RP model in a Python language program. The simulation algorithm of the ions motion calculation is similar to the one presented in references [46, 50]. This program enables one to simulate the cooling process, including the embedded ability to estimate the space charge effect by tube-method, while the electrostatic field has an analytical expression. The simulation of cooled beam transport cannot be done in this program because the electric fields created by the injection and extraction electrodes do not have a known analytical expression.

## 5 Experimental setup: Optics system of SHIRaC

The main sections of an RFQ Cooler beam line are: deceleration and injection section, cooling section, and extraction and re-acceleration section [17]. The design of a good optics system is one of the most important topics. These sections have to ensure an efficient transmission of different types of radioactive ions beam. Their optimizations were done by Simion and it will not be outlined in this paper.

### 5.1 Deceleration and injection

To efficiently cool a typical SPIRAL 2 beam and prevent secondary ionization, the injection energy that will bring the ions to the cooling section should be less than the second ionization energy, so less than 200 eV [5, 35]. Therefore, the relatively high energy of beam should be decelerated using a DC electric field. The deceleration can be done by an injection plate electrode setting at the high voltage with the platform and a grounded electrode setting at the mass (figure 2).

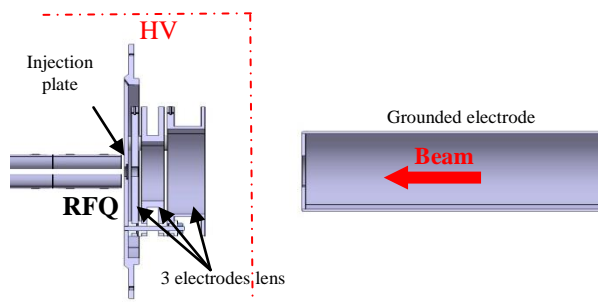

**Fig. 2**: Layout of the deceleration and injection section optics

### 5.2 Cooling section

It consists of the main cooler chamber, setting at HV. It encloses the buffer gas and the RF quadrupole electrodes. It is devoted to trap efficiently the injected beam and to cool it progressively with the buffer gas. To guide the ions along the RF quadrupole up to the extraction cell, the RFQ electrodes are segmented and a DC axial potential is applied on these segments [51, 52]. For RFQ pressure of a few Pascal, the quadrupole length should be in the range of 300-1000 mm [14]. We have chosen a length of 725 mm so 18 segments of 40 mm of length. Depending on the compromises taken in the capture of the large incoming beam, the handling of the high intensity beam and the usage of the higher potential well depth, this can be provided in quadrupole inner radius of 5 mm.

### 5.3 Extraction and reacceleration

Once the ions are cooled we should set up an available optical system to extract them. Due to the diffusion of gas and the space charge effects, these ions must be simultaneously extracted and accelerated because their energies are too low to overcome these effects. The optical design of this section is not easy because the ions motions are in phase with the RF electric field [18]. It is shown in figure 3 that the cooled ions are extracted and accelerated by a DC electric field provided by an extraction plate electrode setting at HV. The three electrodes lens is set up to prevent ions loss.

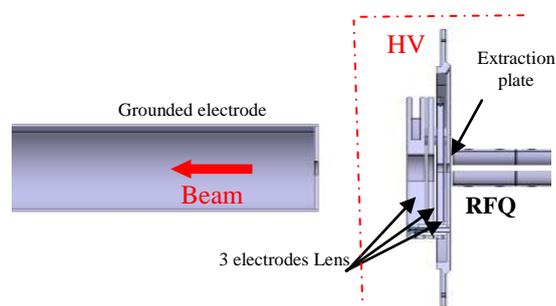

**Fig. 3**: Layout of the extraction and re-acceleration system optics.

## 6 Results and discussion

### 6.1 Simulation of the cooling process

In this part we will discuss and compare the simulations of the cooling process by Simion and python programs.

We will deal the dependence of the transmission efficiency, energy spread ΔE and emittance ε to two effects the RFQ pressure and the space charge.

Our simulations were done with singly charged $_{133}Cs^+$ ions at the following parameters: a typical SPIRAL 2 beam of ε = 80 π.mm.mrad and 140 eV of kinetic energy at the RFQ entrance. The RF electronic system is 4.5 MHz of RF frequency, 5.52 kV of RF amplitude (Mathieu parameter q=0.4). The DC axial guiding field is at 16V/m. About the vacuum system, the distribution of buffer gas pressure is based on the table presented in section 2.2. At both the entrance and cooling section exit, the variation of pressure is assumed to be linear.

### 6.1.1 Pressure dependence

For the RFQ cooler the ions losses can be mainly produced during the deceleration and the cooling section stages. In the first stage, when the ions hit the RF electric phase in the RFQ entrance the peripheral ions can be lost [5]. Along the second stage, the exhaust of ions depends on the confinement capacity which increases with the cooling power (i.e the RFQ pressure), and decreases with the space charge as shown in the equation 3. At a given RFQ pressure, the cooling power is constant along this section but the space charge increases because the ions velocity decreases, see equation 5.2. Therefore, the confinement capacity is reducing along this section and ions losses can be occurred. When the RFQ pressure increases the cooling power also increases. Then, the space charge effect will be compensated by the cooling power and the confinement capacity is improving. Hence, as you can see in the figure 4 an increase in the transmission coupled with a decrease in both ΔE and ε. More than 60 % of transmission can be obtained for RFQ pressure over 2 Pa. In this pressure range, both ΔE and ε become more constant and close to 1eV and 1π.mm.mrad, respectively.

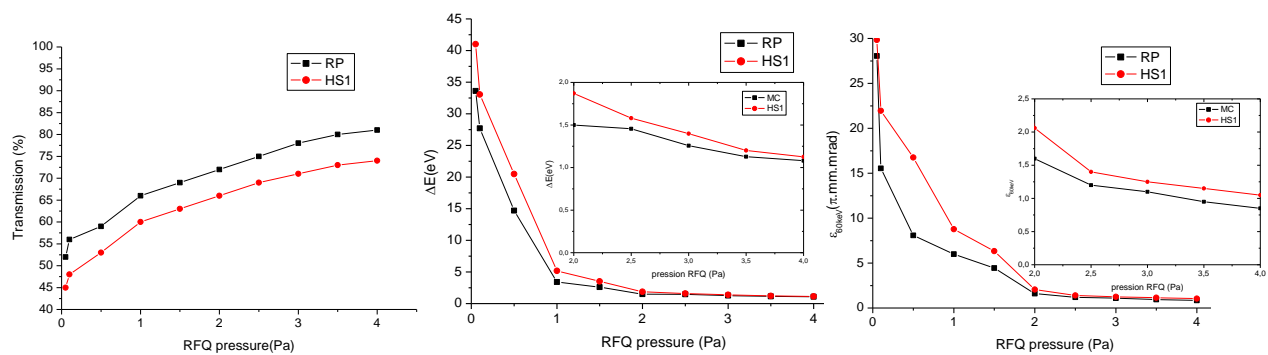

**Fig. 4**: simulation of the RFQ pressure effect on the beam properties: features of the 1 μA cooled $_{133}Cs^+$ ions beam taken at the end of RFQ.

### 6.1.2 Space charge dependence

In this section, we will discuss the space charge effect on the cooled beam properties for 2.5 Pa of RFQ pressure.

At a given RFQ pressure the cooling power is constant. But, along the cooling section the space charge effect increases with the current beam because the ions velocity decreases. Therefore, the degradation of confinement capacity will be significant. This degradation gives rise to an ions number reduction which can be trapped along the cooling section. Hence, the decrease of transmission with current beam, as shown in figure 5: for intensities ranging from 100 nA up to 1μA, the transmission falls slightly by 15 % and it remains more than 60 %.

At the thermal equilibrium, the space-charge effect increases the temperature of cooled ions beam therefore it increases the emittance ε [17, 5]. On the below figure, ε increases slightly as a function of beam current because the potential well depth is 10 times higher than the space charge well depth (D≈550 eV and DSC≈52 eV for 1μA beam of 1 eV of kinetic energy) and it is around 1.25 π.mm.mrad for 1 μA beam.

We can also observe on figure 5 that ΔE undergoes a noticeable variation and its value is around 1.4 eV. As ΔE stands for the spread longitudinal component velocities of ions, so it remains almost constant because the space charge effect acts only on the transversal component of the ions velocities.

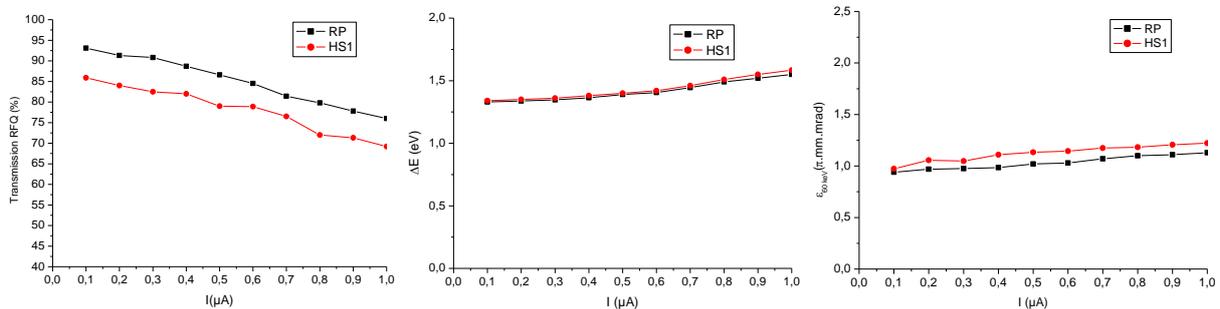

**Fig. 5**: Space charge effect on the characteristics of cooled $_{133}Cs^+$ ions beam.

### 6.1.3 Comparison HS1/RP models

In figures 4 and 5, the simulation results by RP and HS1 model are presented. They are similar in term of spread energy ΔE. Moreover, a difference in terms of transmission and emittance can be seen. We will study quantitatively this difference.

On the one hand, The RP model has a longer interaction range and has about 9 times more collisions than the HS1 model [17, 5]. In the RP model, the collisions change the ions directions greatly and the ions trajectories follow deviate more abruptly than they do in simulations with the HS1 model [17]. Thus with the RP model the power to control the ions is easier than with HS1 model, therefore we can cool more ions with the RP model. Likewise, the cooling process is best performed with the RP model, as the simulation results show it.

On the other hand, the differences in simulation results by RP and HS1 models remain less than 6%. This slight difference entices us to simulate the transport of cooled ions beam with Simion software and the HS1 model.

### 6.2 Simulation of cooled ions beam transport

As presented in the previous paragraph, the cooled beam properties obtained at the RFQ limit are in agreement with the HRS requirements for current beam going up to 1 µA. The main challenge is to maintain the features of these cooled beams up to the extraction section exit. In this part, we will study the degradations of these features at the exit of extraction section and their reasons.

Firstly, the optimum cooling in term of RFQ pressure will be reviewed. Secondly, the degradation of cooled $Cs^+$ ions beam will be studied. Finally, the mass effect on the cooling power will be discussed.

### 6.2.1 Optimum cooling pressure

As the transmission efficiency always being an important feature of RFQ Cooler. The study of optimum cooling will be based on this feature. To do this, we will seek the perfect RFQ pressure value which corresponds to a maximum of transmission.

On the below figure we have the transmissions at the extraction section exit and RFQ limit as a function of the RFQ pressure. Contrary to the behavior of the transmission with the RFQ pressure shown at the RFQ limit, the transmission at the extraction section exit increases from zero, below 0.1 Pa, up to a maximum at 2.5 Pa then it decreases again. The increase of transmission is due to the increase of the cooling power with the RFQ pressure. However, the decrease explains the dominance of the buffer gas diffusion effect, beyond the RFQ limit, compared to the cooling power. The maximum at 2.5 Pa corresponds to an optimum of cooling.

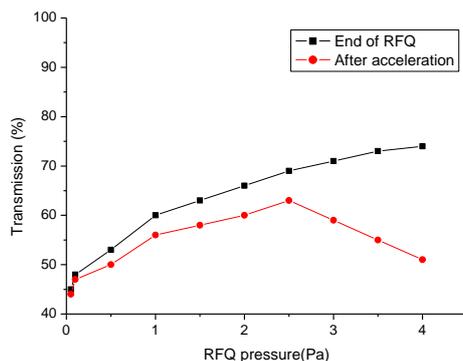

**Fig. 6**: Simulation of the transmission as a function of the RFQ pressure for Cs$^+$ ions beam of 1μA of intensity at the RFQ limit and at the extraction section exit.

### 6.2.2 Cooled beam features after acceleration

The absence of confinement potential beyond the RFQ limit induces us to assess the degradations brought to the cooled beams properties during its transport from the RFQ limit to the extraction cell exit.

Recalls that from the RFQ limit the ions cross a region of length around 5 mm, in which the pressure is close to the RFQ pressure, before being accelerated. As the kinetic energy of ions does not exceed few eV, the degradations of beam by both the space charge and buffer gas diffusion will be very significant. The comparative simulation results of Cs$^+$ cooled beam properties versus the current beam for an RFQ pressure of 2.5 Pa is shown in figure 7. These degradations can be seen on the gap between the beam properties at the RFQ limit and at the extraction section exit.

The evaluation of the contribution of each effect can be done by assuming that: the diffusion of gas contribution is independent to the current beam and its value can be estimated to the degradation of beam at a very low intensity. However, space charge contribution should be increased with the current beam. The first effect contribution can be seen on the energy spread, the emittance and the transmission which are degraded respectively, of 3 eV, 0.25 π.mm.mrad and 5%. The contribution of the second effect is most noticeable on the emittance and the energy spread which increase linearly with the current beam and for 1 μA the beam properties are: ΔE ≈ 5.9 eV, ε ≈ 2.4 π.mm.mrad and transmission around 65 %.

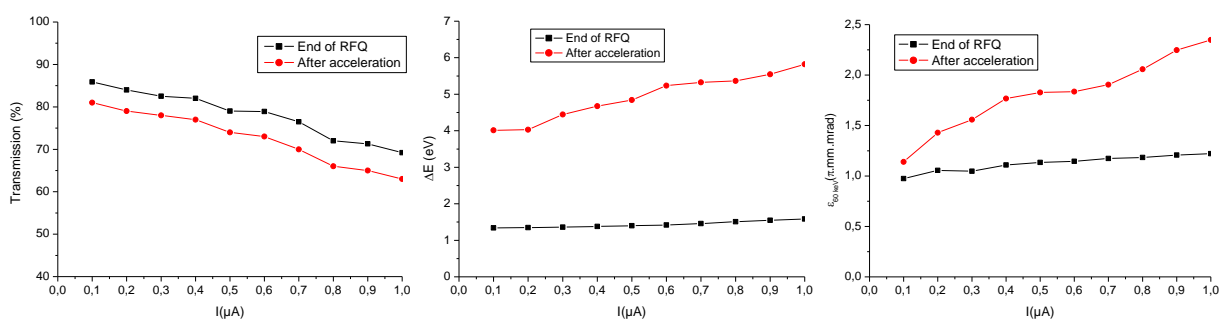

**Fig. 7**: Simulations of the transport of $_{133}$Cs$^+$ cooled ions beam from the RFQ limit: variation of beam properties with current beam at the RFQ limit and at the extraction section exit.

### 6.2.3 Masses effect

Recall that in SHIRaC it will be possible to study the widest range of ions masses, from 10 up to 250 a.m.u. That is why the simulation of ion mass effect on the cooling power should be treated.

At a given RF frequency F and q parameter, such as F=4.5 MHz and q=0.4, the confinement potential well depth $D_{max}$ increases with the ion mass (equation 4 and 1). As the well depth $D_{SC}$ is independent to it (equation 5.1), the decrease of the cooling power appears more clearly along the cooling section when the ion mass decreases. Therefore, degradation of cooled beams properties can be occurred. Beyond the cooling section these cooled beam undergo a second degradation on account of the space charge and the buffer gas diffusion. The table 2 shows these degradations for three mass: heavy mass ($_{133}Cs^+$ ions), medium mass ($_{87}Rb^+$ ions) and light mass ($_{23}Na^+$ ions).

When the ion's mass is close to the buffer gas mass, the RF heating can contribute to degrade the cooled beam. This might explain the most degradation of spread energy in the case of $Na^+$ ions relative to others ions.

| Ion mass | $_{133}Cs^+$ | $_{87}Rb^+$ | $_{23}Na^+$ |
|---|---|---|---|
| $D_{max}$ (eV) | 552 | 361 | 95 |
| Transmission % | 65 | 49 | 25 |
| $\Delta E$ (eV) | 5.9 | 6.1 | 7.7 |
| $\varepsilon$ ($\pi$.mm.mrad) | 2.4 | 2.7 | 3.5 |

**Table.2**: Masses effect on the cooling power: features of 1 µA cooled ions beam at the extraction section exit.

Despite these degradations, the beam properties uphold the HRS requirements in terms of emittance and spread energy. Only the spread energy is far to the HRS requirement.

### 6.3 Experimental results: Transmission

This section presents the first experimental results in term of transmission for the $_{133}Cs^+$ ions beam which is provided by an IGS4 ionization surface source.

The measurement of transmission is based on two Faraday cups placed on either side of RFQ section: one is placed at the injection section entrance and the second is placed at the extraction section exit.

Firstly, the space charge effect on the transmission efficiency of ions will be studied. On figure 8 both experimental and simulated transmission results according to current beam are shown. This effect is clear on the transmission which decreases for current beam going from a few nA up to 1 µA. This decrease happens gradually because along the cooling section the confinement potential well is very deep and beyond the RFQ limit both the space charge and the buffer gas diffusion are low on the transmission. As it is shown in simulation, the transmission remains above 60% for intensities going up to 1 µA.

The slight difference between the experimental and simulations results presented on figure 8 is explained by the unrealistic physics model of collision. Experimental results match simulation ones.

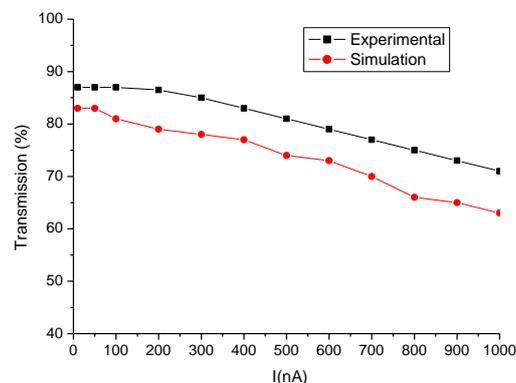

**Fig. 8**: Experimental and simulated results of space charge effect on the transmission for 1 µA $Cs^+$ ions beam.

To understand the RF voltage effect on the beam properties it is convenient to study the q parameter versus beam properties, such as the transmission. On the figure below, we present the transmission efficiency versus the q parameter for 1 µA ions beam. We can see that the optimum cooling is obtained with 2.5 Pa of RFQ pressure. With this pressure, the transmission is around 71 % for q=0.4. For q>0.6 the transmission decreases rapidly because the RF heating can occur.

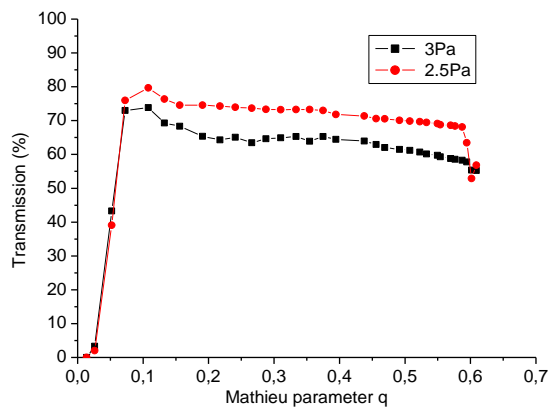

**Fig. 9**: Transmission efficiency for continuous $_{133}Cs^+$ ions beam as a function of q parameter for 1 µA current beam.

## 7  Conclusion and outlook

Both simulation and experiment show close results for intensities never handled so far. The SHIRaC prototype achieves transmissions efficiency over 60%. However, the validation of this project would require the energy spread and the emittance measurements. Nevertheless these studies will be presented in a next paper. Results presented in this paper are parts of my PhD thesis work.

Based on this study the effects which act most on the cooled beam are the space charge and the diffusion of buffer gas in the region beyond the RFQ limit. In order to avoid them, the RF confinement potential should be large enough in this region. This can be done in several ways. One such solution is to change the extraction electrodes design. Another option is to set-up a miniature RFQ going from the last electrode of the RFQ to the extraction cell [22, 53]. The last option is being developed and will be tested later on.

### Acknowledgments

We thank Pr G.Ban, D.Durand and J.C.Steckmeyer for their advices during my PhD thesis and the LPC Caen instrumentation team for their assistance with the development of the SHIRaC beam line as well as Jean-Francois Cam et Christophe Vandamme for their contributions in the development of the RF system and the vacuum system.